\documentclass{article}
\usepackage{spconf,amsmath,graphicx}
\usepackage{verbatim}
\usepackage{xcolor}
\usepackage{graphicx}
\usepackage{multirow}
\usepackage{booktabs}
\usepackage{cite}

\title{Bi-APC: Bidirectional autoregressive predictive coding for unsupervised pre-training and its application to children's ASR}
\name{Ruchao Fan, Amber Afshan and Abeer Alwan\thanks{This work was supported in part by National Science Foundation (NSF) Grant \#1734380.}}
\address{Department of Electrical and Computer Engineering, University of California Los Angeles, USA}
\begin{document}
%
\maketitle
\begin{abstract}
We present a bidirectional unsupervised model pre-training (UPT) method and apply it to children’s automatic speech recognition (ASR). An obstacle to improving child ASR is the scarcity of child speech databases. A common approach to alleviate this problem is model pre-training using data from adult speech.
Pre-training can be done using supervised (SPT) or unsupervised methods, depending on the availability of annotations. Typically, SPT performs better. In this paper, we focus on UPT to address the situations when pre-training data are unlabeled. Autoregressive predictive coding (APC), a UPT method, predicts frames from only one direction, limiting its use to uni-directional pre-training. Conventional bidirectional UPT methods, however, predict only a small portion of frames. To extend the benefits of APC to bi-directional pre-training, Bi-APC is proposed. We then use adaptation techniques to transfer knowledge learned from adult speech (using the Librispeech corpus) to child speech (OGI Kids corpus). LSTM-based hybrid systems are investigated. For the uni-LSTM structure, APC obtains similar WER improvements to SPT over the baseline. When applied to BLSTM, however, APC is not as competitive as SPT, but our proposed Bi-APC has comparable improvements to SPT.

\end{abstract}

\section{Introduction}
\label{sec:introduction}

One of the challenges faced in developing automated and individualized educational and assessment tools for children is the performance lag in child ASR compared to adult ASR~\cite{kennedy2017child}. Challenges arise, in part, from difficulties in acoustic and language modeling of child speech. Due to different growth patterns of children and motor control issues, children's speech has a higher degree of intra-speaker and inter-speaker acoustic variability~\cite{lee1999acoustics}. Additionally, children's speech is characterized by significant mispronunciations and disfluencies~\cite{shivakumar2014improving}. Another challenge is the lack of publicly-available child speech databases. Interestingly, with enough training data, the performance of child ASR using CLDNN-based hybrid models was shown to be comparable to adult systems~\cite{liao2015large}. To alleviate the data scarcity problems, data-efficient TDNN-F network for child ASR was proposed in~\cite{wu2019advances}.    


Model pre-training with a data-sufficient task is another successful approach to address the data scarcity issue. When combined with fine-tuning, model pre-training can transfer the knowledge learned from one task to another~\cite{dong2019unified}. Supervised pre-training (SPT) has been effectively applied to cross-lingual~\cite{huang2013cross} and child ASR~\cite{shivakumar2020transfer,gale2019improving, tong2017transfer}. However, obtaining transcriptions is not always feasible. Recently unsupervised representation learning was proposed for situations when transcriptions are not available. This approach could be used for--(i) feature extraction, and (ii) model initialization, referred to as unsupervised pre-training (UPT). Common unsupervised techniques used as feature extractors include autoregressive predictive coding (APC)~\cite{Chung2019,ravi2020exploring} and contrastive predictive coding (CPC)~\cite{oord2018representation}. APC predicts a future frame from previous ones to learn speech representation while CPC considers samples randomly selected from the waveform, referred to as "negative samples". Most UPT methods apply BERT-style pre-training mechanisms, which reconstruct the masked frames (frames masked to zero as input) from unmasked frames using bidirectional information~\cite{jiang2019improving, liu2020mockingjay, song2020speech, baevski2019effectiveness, wang2020unsupervised}. However, UPT methods have not been used for child ASR. 

Unlike APC, most UPT methods mask only partial frames for prediction limiting the pre-training model from learning a more comprehensive representation. APC, which is mostly used for feature extraction, is constrained to learning from only one direction, limiting its use in bi-directional sequential models. Bi-directional models provide better performance for ASR systems in comparison to their uni-directional counterparts~\cite{zeyer2017comprehensive}. To fully exploit the potential of APC for bidirectional models, we propose a novel bidirectional APC to use as an UPT and we refer to this technique as Bi-APC. 


We evaluate supervised and unsupervised pre-training methods and investigate their ability of transferring knowledge learned from adult speech to child speech in the context of LSTM-based acoustic models. We also evaluate our proposed Bi-APC technique against conventional bidirectional pre-training methods such as MPC. The remainder of the paper is organized as follows. Section 2 presents the proposed Bi-APC technique along with SPT and APC. Section 3 describes the experimental setup, followed by results and discussion in Section 4. Section 5 concludes the paper. 


\section{Model Pre-training Methods}
\label{sec:methods}
Model pre-training learns common knowledge from a data-sufficient task and then transfers the knowledge learned to a low-resource task. In this paper, we aim to transfer the knowledge learned from adult speech to child speech. We use the pre-training methods described in this section for adult model training. Long short-term memory (LSTM) based networks are chosen as acoustic models, which are then used to form a hybrid HMM-LSTM ASR system. Based on the training mechanism, we can summarize the pre-training methods into two categories--supervised and unsupervised.

\subsection{Supervised Pre-training}
\label{ssec:spt}
Recently, supervised pre-training has been successfully used in child ASR~\cite{shivakumar2020transfer} and is frequently referred to as transfer learning. Specifically, suppose the output of the LSTM is $Y=\{y_1, y_2,\dots,y_T\}$ and the corresponding frame-level label obtained from forced alignment is $\hat{Y}=\{\hat{y_1},\hat{y_2},\dots,\hat{y_T}\}$, the supervised training aims to optimize the cross-entropy loss function:
\begin{equation}
  L_{\text{CE}} = -\sum_{t=1}^T\sum_{c=1}^C\hat{y_t^c}log(y_t^c)
  \label{eq:celoss}
\end{equation}
where $C$ is the number of output categories (HMM states). The parameters in the LSTM are then utilized as the initialization for child acoustic model training except for the last feed-forward layer due to the different state space between adult and child models.


\subsection{Unsupervised Pre-training}
Different from supervised pre-training, unsupervised pre-training does not require speech labels. Most of the unsupervised pre-training methods use either prediction or mask and reconstruction, where the supervision is the speech signal itself. In this section, we first review the APC for uni-LSTM pre-training and then show how we can extend the APC to bidirectional LSTM (BLSTM) pre-training.

\subsubsection{Autoregressive Predictive Coding (APC)}
\label{sssec:apc}
APC utilizes the shifted input sequence as supervision and tries to predict the frame $n$ steps ahead of the current frame with information from previous frames. As it is a regression-based prediction task, we consider the $L_1$ distance. Suppose the input feature sequence is $X=\{x_1,x_2,\dots,x_T\}$, then the pre-training model is trained with the following loss function:
\begin{equation}
  L_{\text{APC}} = \sum_{t=1}^{T-n}(|x_{t+n}-y_t|)
  \label{eq:apcloss}
\end{equation}
where $n$ is a fixed value as a hyper-parameter. A key difference from~\cite{Chung2019} in the usage of APC in this paper is that we utilize the pre-training model for parameter initialization instead of feature extraction. 
The reason is that we do not expect APC training with adult data as a feature extractor to result in improvements for child ASR, due to the large acoustic mismatch between adult and child speech. Nevertheless, the mechanism of APC can be used for LSTM pre-training from only one direction, and thus does not fully exploit information from both directions.

\begin{figure}[t]
  \centering
  \includegraphics[width=8cm,height=4.5cm]{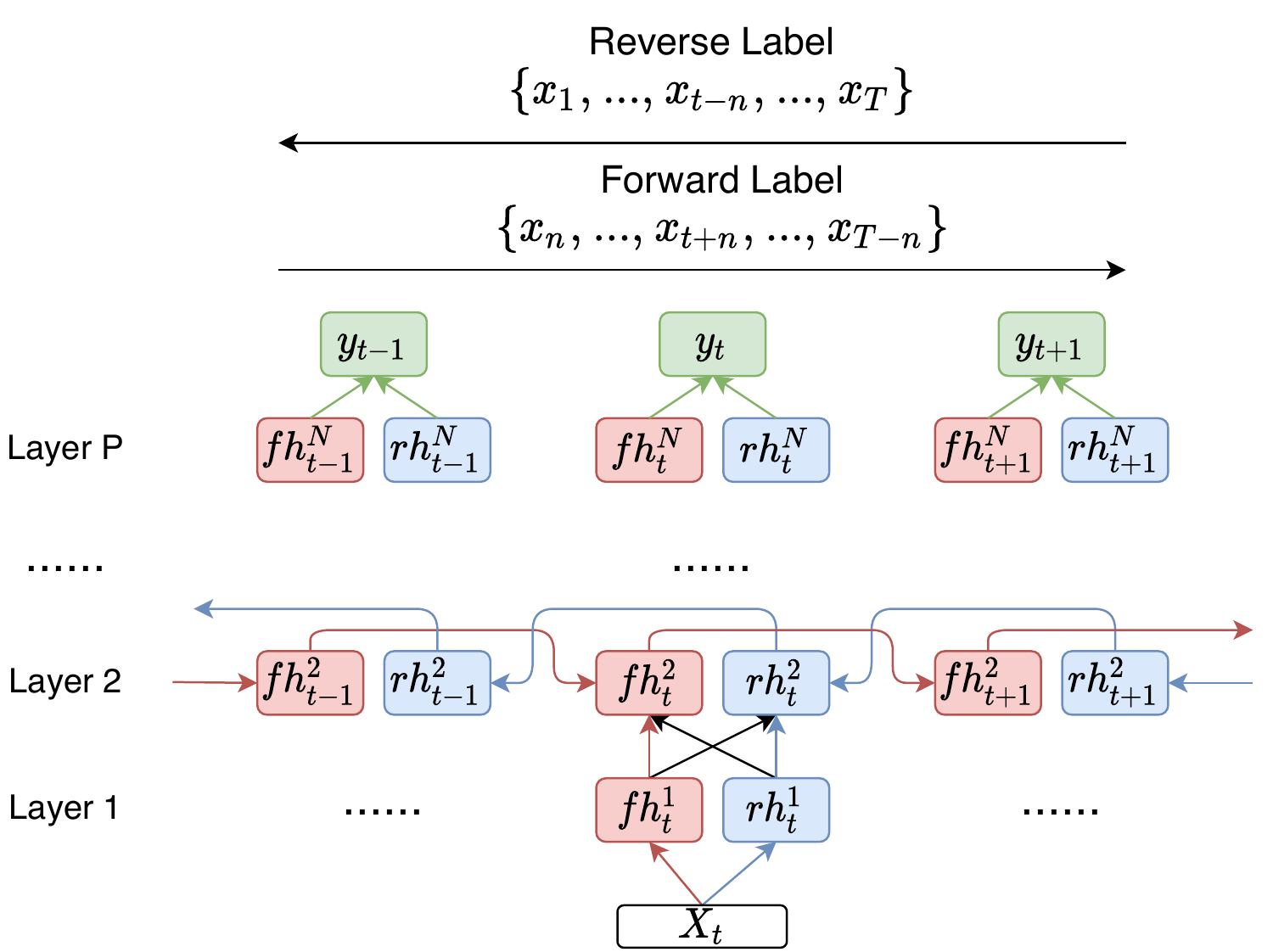}
  \caption{Illustration of Bi-APC pre-training for BLSTM. Red parts and blue parts are the forward-related and reversed-related parameters and computations, respectively. $fh$ and $rh$ indicate the hidden states of the forward and reversed calculations, respectively, at each layer.}
  \label{fig:bapc}
\end{figure}

\subsubsection{Bi-APC: Extending APC to learn from both directions} 
\label{sssec:bi-apc}
The mechanism of APC is well suited for uni-directional structures such as uni-LSTM. However, BLSTMs usually provide better performance than uni-LSTMs as they learn from both directions. Therefore, we propose a bidirectional APC (Bi-APC), which extends APC to exploit its potential for BLSTM pre-training. The idea of Bi-APC is to add a reversed version of APC prediction, where we predict the frame $n$ steps behind the current frame given all future frames.

Figure~\ref{fig:bapc} shows how to use Bi-APC for BLSTM pre-training. To prevent equivalent mapping in the network, the outputs of the BLSTM should not contain information about the corresponding supervisions. 
We, therefore, split the BLSTM into forward-related and reverse-related parts as shown in red parts and blue parts in Fig.~\ref{fig:bapc}, respectively, including the parameters (arrows) and outputs (rectangles) at each layer. When computing the outputs $Y^{fwd}=\{y_1^{fwd},y_2^{fwd},\dots,y_T^{fwd}\}$ in the forward direction, the values of the blue rectangles are set to zero to exclude the information that are extracted from the frames on the right side. The reversed-related parameters are also not updated. The same strategies are used in the computation of outputs $Y^{rev}=\{y_1^{rev},y_2^{rev},\dots,y_T^{rev}\}$ in the reversed direction. The parameters in black arrows are not trained in the pre-training since they allow for an illegal information exchange from different directions. The green arrows are the shared parameters which are not used in the fine-tuning. The BLSTM is then pre-trained by optimizing the APC from both directions as:

\begin{equation}
  L_{\text{Bi-APC}} = 0.5 \cdot \sum_{t=1}^{T-n}|x_{t+n}-y_t^{fwd}| + 0.5 \cdot \sum_{t=n+1}^{T}|x_{t-n}-y_t^{rev}|
  \label{eq:bapcloss}
\end{equation}
where task ratios are set to 0.5 as both directions have the same importance. Note that we can also train an APC with uni-LSTM and only initialize the parameters of the red parts in Figure~\ref{fig:bapc}. We still denote this pre-training as APC in the experimental results.

\section{Experimental Setup}
Experiments were conducted using Kaldi~\cite{povey2011kaldi} and Pykaldi2 \cite{lu2019pykaldi2}. Pykaldi2 is used to train the neural networks for the hybrid system and Kaldi is used for WFST-based decoding.

\subsection{Database}
\label{ssec:database}
For the pre-training task, Librispeech~\cite{panayotov2015librispeech} was used because it is the largest publicly-available adult speech corpus and is mainly read speech. The test set of the Librispeech corpus is split into “clean” and “other” based on the quality of the recorded utterances, where the ”other” refers to noisy data, and are used to evaluate the adult ASR system. 

For the fine-tuning experiments, the scripted part of the OGI Kids' Speech Corpus~\cite{shobaki2000ogi} was used. It contains speech from approximately 100 speakers per grade saying single words, sentences and digit strings. The utterances were randomly split into training and test sets without speaker overlap, where utterances from 30\% of the speakers were chosen as the testing data, denoted as ogi-test. As a result, nearly 50 hours of child data were used to train the child ASR system. 

\subsection{Acoustic Model Setup}
\label{ssec:pre-train setup}

The initial experiments used GMM model training. The Librispeech recipe in kaldi was used for pre-training and the JHU OGI recipe~\cite{wu2019advances} was applied for fine-tuning. The GMM models were then used to obtain the frame-level alignment for DNN-based acoustic model training. The HMM states were 5776 and 1360 for adult and child models, respectively.

Uni-LSTM and BLSTM were chosen as acoustic models to compare pre-training methods. 80-dimensional Mel-filter bank features (which is common for UPT) were extracted from each 25ms window with a 10ms frame shift as the input. No frame stacking or skipping was applied. Hence, the output dimension for the unsupervised pre-training task is 80. The uni-LSTM model consists of 4 uni-LSTM layers with 800 hidden units, while the BLSTM model has 4 BLSTM layers with 512 hidden units in each direction. Batch normalization and dropout layers with a 0.2 dropout rate were applied after each LSTM layer. The output of the LSTMs were then transferred into either the state space for classification or the feature space for prediction with a single feed-forward layer.

All models were trained with a multi-step schedule, where the learning rate was held in the first 2 epochs and then was exponentially decayed to a ratio $\lambda$ of the initial learning rate in the remaining epochs. For pre-training tasks, 8 epochs were used with the initial learning rate of $0.001$ and $\lambda=0.1$. For the fine-tuning tasks, we trained the models with 15 epochs. The learning rate starts from 2e-4 to 2e-6. The last three model checkpoints were averaged as the final model for evaluation. For both APC and Bi-APC training, the time shift $n$ was heuristically set to 2. Sequence discriminative training was not applied in our experiments since our goal is to compare different pre-training methods.

\subsection{Language Model Setup}
All experiments use the same lexicon and language models from the original Librispeech corpus. Specifically, the 14M tri-gram (tgsmall) language model was used for first pass decoding, and the 725M tri-gram (tglarge) language model was used for rescoring. We report the results of rescoring.

\begin{table}[t]
  \caption{WERs of baseline systems, including uni-LSTM and BLSTM trained with Librispeech and OGI data, respectively.}
  \label{tab:baseline}
  \small
  \centering
  \begin{tabular}{l c c c}
    \toprule
    \multirow{2}{*}{WERs(\%)} &\multicolumn{2}{c}{Libri-adult} & \multicolumn{1}{c}{Children} \\
    \cmidrule{2-4}
     ~ & test-clean & test-other & ogi-test  \\
    \midrule \midrule
    \multicolumn{4}{c}{Adult Model - Librispeech} \\
    \midrule
    uni-LSTM & 5.71 & 15.15 & 65.90 \\
    BLSTM & 4.90 & 12.59 & 59.12  \\
    \midrule
    \multicolumn{4}{c}{Child Model - OGI Corpus} \\
    \midrule
    TDNN-F \cite{wu2019advances} & - & - & 10.71  \\
    uni-LSTM  & 95.77 & 97.28 & 12.58    \\
    BLSTM   & 86.82 & 92.15 & 9.16       \\
    \bottomrule
  \end{tabular}
\end{table}

 \begin{table}[th]
   \caption{Performance comparison of supervised pre-training (SPT) and unsupervised pre-training (UPT) in terms of WER (\%) for both LSTM and BLSTM acoustic model architecture. The results are for ogi-test. We also provide word error rate reduction (WERR) compared to the baseline.}
   \label{tab:supservised and unsupervised}
   \centering
   \small
   \begin{tabular}{c|c c c c c}
   \toprule
   \multicolumn{2}{c}{WERs(\%)} & uni-LSTM & WERR & BLSTM & WERR \\
     \hline \hline
     \multicolumn{2}{c}{Baseline}  & 12.58 & - &  9.16  & -   \\
     \multicolumn{2}{c}{SPT}  & 11.85 & 5.8\% & 8.46 &  7.6\%   \\
     \hline
     \multirow{3}{*}{UPT} & MPC \cite{jiang2019improving} & - & - & 9.02 & 1.5\%   \\
     ~ & APC & 11.76 & 6.5\% & 8.85 & 3.4\% \\
     ~ & Bi-APC  & - & - &  8.57 & 6.5\% \\
   \bottomrule
   \end{tabular}
 \end{table}

\section{Results and Discussion}
\subsection{Baseline}

We first show the results of the baseline models in Table~\ref{tab:baseline}. Here we compared two models--(a) adult model trained using Librispeech and (b) child model trained using the OGI speech corpus. We evaluated these models on test-clean and test-other from Librispeech and also on the OGI test. We compared uni-LSTM and BLSTM acoustic model architectures for both setups. For the adult model, we obtained performances similar to previously published results~\cite{panayotov2015librispeech}. Adult models were also used to test on ogi-test that has an acoustic domain mismatch resulting in high WERs for LSTM models. 

For child models, the performance on Librispeech degrades drastically with both uni-LSTM and BLSTM models. To compare with existing results in the literature, we evaluated the TDNN-F acoustic model trained with the OGI corpus~\cite{wu2019advances}. We see that the uni-LSTM performed worse than TDNN-F but BLSTM outperformed TDNN-F, thus motivating us to explore model pre-training for the BLSTM system.

\subsection{Comparison of Pre-training Methods for Child ASR}
This paper aims at exploring the performance of supervised (SPT) and unsupervised pre-training (UPT) for children's ASR. As mentioned in Section~\ref{ssec:database}, we used Librispeech for pre-training and OGI for fine-tuning the model. Table~\ref{tab:supservised and unsupervised} presents results of fine-tuning on both uni-LSTM and BLSTM architectures, evaluated on the OGI test. Note that, different from~\cite{shivakumar2020transfer}, all layers were updated during fine-tuning since this was the best setting for our experiments.
 
Table~\ref{tab:supservised and unsupervised} shows that SPT improved the performance of the uni-LSTM model to 11.85\% and was better than the baseline without pre-training. Interestingly, unsupervised pre-training using APC also provides improvement (11.76\%) similar to that of SPT with the uni-LSTM model. 

As mentioned earlier, BLSTM has better performance than uni-LSTM. SPT resulted in the best performance (WER, 8.46\%) among the pre-training methods applied to BLSTM. Note that, to perform UPT, we first used APC to pre-train only the forward path parameters of BLSTM, resulting in a WER of 8.85\%. We then compared it to a widely used bidirectional pre-training method, the masked predictive coding (MPC)~\cite{jiang2019improving} and showed that MPC (9.02\%) performed worse than APC (8.85\%). We assume the reason is that MPC has fewer frames to be predicted (only 15\% of the frames were randomly masked) although MPC can learn from both directions. The proposed Bi-APC achieved a WER of 8.57\% that is comparable to the SPT. This can be valuable when there is a large amount of data without transcriptions. Since the pre-training task is a 960-hour dataset, UPT could possibly benefit from more unlabeled data. Recent works have shown that self-attention layers are better for acoustic modeling than the BLSTM \cite{wang2020transformer,lu2020exploring}. It will be interesting to see how Bi-APC could be extended to other model topologies, which is an important issue for future research.

\begin{table}[t]
  \caption{BLSTM-based ASR performance breakdown based on age groups of kindergarten to grade 2, grade 3-6 and grade 7-10.}
  \label{tab:age comps}
  \centering
  \begin{tabular}{l c c c}
  \hline
  \multirow{1}{*}{WERs(\%)} & K0-G2 & G3-G6 & G7-G10 \\
    \hline \hline
    Baseline & 18.87	& 7.24 &	5.51     \\
    +SPT & 17.43 &	6.66 &	5.11       \\
    +APC & 18.07 & 7.03 & 5.40 \\
    +Bi-APC  & 17.23 & 6.91 & 5.26      \\
     
  \hline
  \end{tabular}
\end{table}

\subsection{Performance Breakdown based on Age Groups}

To obtain an insight into the influence of the speaker's age on the performance of pre-training methods, in Table~\ref{tab:age comps}, we present results based on age groups in the OGI dataset. Similar to~\cite{safavi2016speaker}, three different age groups were selected--kindergarten to grade 2, grade 3-6, and grade 7-10. We present the results using the BLSTM model. For younger children (kindergarten- grade 2), the Bi-APC provided slightly better results compared to SPT. In contrast, we did not observe any such improvement in the older age groups for children. This trend could mean that UPT may be capturing a representation crucial to the performance of very young child speech, whose speech is more variable and difficult to recognize than older children~\cite{yeung2018difficulties}. Further research is required to explore the usage of the approach more effectively for children's ASR.
\section{Conclusions}
\label{sec:conclusion}

In this paper, we proposed a bidirectional pre-training (Bi-APC) method. We also compared supervised and unsupervised model pre-training methods for child ASR. We showed that standard APC could be well applied to uni-LSTM pre-training, achieving about 6.5\% relative WER improvement over the uni-LSTM baseline without pre-training. However, APC lost its superiority when applied to the BLSTM structure and had a performance gap with SPT. Our proposed Bi-APC addressed these issues and resulted in comparable performance to SPT. We further analyzed the performance of child speech for different age groups. Results showed the potential of unsupervised pre-training for younger child speech, and we achieved the best-reported ASR result (a WER of 8.46\% for SPT) for the OGI Kids corpus. The proposed Bi-APC achieved a WER of 8.57\%, performing better than other UPT methods such as APC and MPC.

\vfill\pagebreak


\bibliographystyle{IEEEbib}
\bibliography{refs}
\end{document}